\documentclass[sigconf,authorversion,nonacm]{acmart}
\AtBeginDocument{%
  }

\acmConference[CHI '25: Generative AI and Accessibility Workshop]{Generative AI and Accessibility Workshop}{April 27,
  2025}{Yokohama, JP}

\begin{document}


\title[Generative AI-based Exposure Visualization for Therapy of CPTSD]{Describe Me Something You Do Not Remember - Challenges and Risks of Exposure Design Using Generative Artificial Intelligence for Therapy of Complex Post-traumatic Stress Disorder}

\author{Annalisa Degenhard}
\affiliation{%
    \institution{Institute of Media Informatics, Ulm University}
  \city{Ulm}
  \country{Germany}
}
\email{annalisa.degenhard@uni-ulm.de}

\author{Stefan Tschöke}
\affiliation{%
    \institution{Center for Psychiatry Südwürttemberg, Ulm University}
  \city{Ulm}
  \country{Germany}
}
\email{stefan.tschoeke@uni-ulm.de}

\author{Michael Rietzler}
\affiliation{%
    \institution{Institute of Media Informatics, Ulm University}
  \city{Ulm}
  \country{Germany}
}
\email{michael.rietzler@uni-ulm.de}

\author{Enrico Rukzio}
\affiliation{%
    \institution{Institute of Media Informatics, Ulm University}
  \city{Ulm}
  \country{Germany}
}
\email{enrico.rukzio@uni-ulm.de}

\renewcommand{\shortauthors}{Degenhard et al.}

\begin{abstract}
    Post-traumatic stress disorder (PTSD) is associated with sudden and uncontrollable, intense flashbacks of traumatic memories. One form of treatment that has been shown to be effective in reducing the severity of trauma-related symptoms is trauma exposure psychotherapy. It involves the controlled recall of traumatic memories to train coping with trauma flashbacks and to allow autobiographical integration of trauma-associated distressing memories. In particular, exposure to visualizations of such memories is an effective approach for successful recall. However, although this form of treatment has been shown to be effective for various types of trauma, it is still only available for a few types of trauma. This is due to a lack of economic solutions for creating individualized exposure visualizations. This is especially true for the treatment of Complex PTSD (CPTSD), as the traumatic memories of this group are mostly individual, and there are no generic visualizations that work for multiple patients. Generative Artificial Intelligence (GAI) offers a flexible, cost-effective solution. GAI not only provides the novel ability to create individualized exposure visualizations during the therapy session, but also allows patients to participate in the visualization process for the first time. With the advent of novel therapeutic perspectives and the potential to improve access to trauma therapy, especially for CPTSD, the use of GAI also presents challenges and exposes the patient to serious risks. The exceptionally high levels of uncertainty and lack of control that characterize both CPTSD and GAI challenge the feasibility of such solutions. To enable safe and effective three-way communication, it is critical to thoroughly understand the roles of the patient, system, and therapist in the context of exposure visualization and how each can contribute to safety. This paper summarizes perspectives, challenges, and risks for the use of GAI in trauma therapy with a focus on CPTSD.
\end{abstract}


\received{12 March 2025}

\maketitle

\section{Trauma and Dissociation}
If I were to ask you to describe an incomplete memory in such a way that it could be visualized in a medium unfamiliar to you, and that the aspects most relevant to you could be presented in as much detail as possible, would you be able to do so in an instant?
This is one of the main challenges of visualization-based exposure therapy, like Virtual Reality Exposure Therapy (VRET), for Posttraumatic Stress Disorder (PTSD) ~\cite{javidiPosttraumaticStressDisorder2012,maercker2018icd11}. PTSD is a common health problem often characterized by heightened physiological arousal and constant vigilance, uncontrolled intense flashbacks of traumatic memories, and avoidance of situations that may include potential triggers of such flashbacks ~\cite{shalev2024neurobiology,brewin2025post,greenbergPTSD,javidi2012international,javidiPosttraumaticStressDisorder2012}. A common symptom associated with psychological trauma, particularly recurrent interpersonal trauma during childhood, is dissociation~\cite{stolovyDissociationAdjustmentDistress2015}. Dissociation can be roughly described as a coping strategy of the brain that limits the accessibility of cognitive aspects in order to avoid overwhelming experiences~\cite{chuDissociativeSymptomsRelation1990}. This may include discontinuity of consciousness, memory, and perception~\cite{DiagnosticStatisticalManual2013}.

Trauma exposure psychotherapy is currently the most promising form of treatment for PTSD following exposure to single and multiple traumatic events~\cite{HOPPEN2024112}. It focuses on the retrieval of avoided memories in order to train the handling of them~\cite{brewin2025post,cloitre2011treatment} and to allow an autobiographical integration~\cite{berntsen2007trauma,smeets2010Autobiographical}. Recall of traumatic memories may be difficult, depending on the patient's level of dissociation. Approaches include collecting retrospective accounts from traumatized individuals, such as the narrative lifeline approach~\cite{neunerNarrativeExpositionstherapieNET2021}, post-hoc observations, and provoking and then studying traumatic memories and flashbacks in individuals suffering from PTSD. The retrieval process typically takes time, as traumatic memories are initially accessible mostly through flashbacks or intuitive feelings, and individuals must learn to access them consciously~\cite{vanderkolkDissociationFragmentaryNature1995}.

However, the effectiveness of exposure therapy depends on successful confrontation with a traumatic stimulus~\cite{bohus2024dialektisch}. This is often challenged by the high levels of avoidance associated with PTSD~\cite{charltonWaysCopingPsychological1996}. In particular, approaches that rely on patients imagining the traumatic stimulus may be ineffective because therapists cannot control whether the patient is actually imagining what is being pretended~\cite{vincelli1999imagination,difedeInnovativeUseVirtual2002}. 
One solution is provided by visualization-based approaches such as VRET~\cite{meggelenComputerBasedInterventionElements2019,meggelenRandomizedControlledTrial2022}. Visualization of traumatic stimuli can at least partially circumvent avoidance. However, this form of treatment depends on the technical and financial feasibility of visualizing the patient's traumatic stimuli. VRET has shown great feasibility for combat-related trauma, as traumatic memories have been shown to be relatively similar across individuals, and visualizations have been shown to be effective for multiple individuals~\cite{rizzo2021fromcombat}.

Complex posttraumatic stress disorder (CPTSD) can result from exposure to multiple traumatic events~\cite{wilsonPTSDComplexPTSD2004,cloitreDistinguishingPTSDComplex2014}. Unlike monotraumatized individuals, the traumatic stimuli described by affected individuals are often highly individualized, and generalization across patients is difficult~\cite{wilsonPTSDComplexPTSD2004}. CPTSD is also associated with greater levels of dissociation, which makes the recall process more difficult and often follows the principle of trial and error~\cite{brewin2025post,hyland2020relationship}. A common cause of CPTSD is childhood trauma.

\section{Accessibility of Treatment for CPTSD}
 Visualization-based trauma exposure has already been shown to be an effective therapy for childhood trauma in both mono- and complexly traumatized patients using individualized exposure therapy~\cite{chuDissociativeSymptomsRelation1990,HOPPEN2024112}. Psychopharmacological treatment with medication is not currently available~\cite{leichsenring2024Borderline}. Given the prevalence of CPTSD \cite{maercker2022complex,karatzias2017ptsd} and the current lack of other effective forms of treatment, psychotherapy with trauma exposure still seems to be the most promising perspective. However, solutions are needed that allow for a customizable and economical visualization of traumatic memories to improve accessibility~\cite{Nester29072022}. Enabling flexible, economic visualization, generative artificial intelligence (GAI) seems to provide a solution. Through exceptional visualization speed, for the first time, the creation of individualized exposure visualizations within the therapy session is enabled. This could solve the problem of describing an unclear memory when not knowing what is relevant for the system to successfully visualize a memory. As an iterative design process becomes feasible, patients may actively participate in the visualization process for the first time, which may improve the efficacy of a memory-oriented therapy approach.

\section{CPTSD-Related Risks \& Challenges in Exposure Therapy}
    The feasibility of GAI-based exposure visualization in the context of psychotherapy for CPTSD arguably faces a number of risks and challenges on the patient side. These include:\\
    
    \begin{itemize}
        \item[\textbf{R1}] \emph{Intense recall}: Regaining access to previously avoided traumatic memories often causes intense emotions and ego disorganization. This may include severe anxiety, feelings of despair, and impulses toward self-destructive behavior~\cite{herman1987recovery}. Panic attacks may also be triggered~\cite{lyssenkoDissociationPsychiatricDisorders2018}. The result of such intense experiences may be a loss of control over the situation due to absorption~\cite{murrayAbsorptionDissociationLocus2007,glicksohnExplorationsVirtualReality1997,banosPsychologicalVariablesReality1999}.
        \item[\textbf{R2}] \emph{Unintentional Trauma Flashbacks}: Individuals suffering from CPTSD are at exceptionally high risk of being unintentionally triggered, both in terms of timing and intensity. Too much exposure can result in \emph{hyperarousal}~\cite{fordTraumaMemoryProcessing2018}. 
        \item[\textbf{R3}] \emph{Avoidance}: Individuals often tend to avoid confrontation with traumatic memories. This often leads to disengagement in the memory retrieval process, self-disclosure, and dropout from therapy in the worst case~\cite{charltonWaysCopingPsychological1996}.
        \item[\textbf{R4}]\emph{Retraumatization}~\cite{duckworth2012retraumatization}: During traumatic memory recall, patients may attempt to rid themselves of the physically and emotionally draining experiences. This can cause patients to become overwhelmed again, and the exposure becomes retraumatizing rather than healing~\cite{chu1992therapeuticRollerCoaster}.
        \item[\textbf{R5}]\emph{Acceptance}~\cite{kimDevelopmentHealthInformation2012,ICHE6R21997}: Acceptance of the therapy and the technology involved can be a critical factor in the success of the therapy. Low acceptance can lead to disengagement and dropout. 

        \item[\textbf{R6}]\emph{Distortion of reality}~\cite{zoellner2000trauma}: The false belief of remembering something due to a bias during traumatic memory exploration.
    \end{itemize}    

On the one hand, the goals of VR exposure are to reinforce the memory of traumatic events and emotional activation, which also represents a danger if uncontrolled, and the challenge, therefore, lies in the controlled, dosed activation of memory and emotions.

\section{Generative AI-based Visualization}
GAI has made remarkable progress in synthesizing 2D and 3D visualizations, driven by advances in deep learning architectures such as generative adversarial networks, variational autoencoders, and diffusion models. Systems such as DALL-E\footnote {https://openai.com/index/dall-e-3/}, Stable Diffusion\footnote {https://stablediffusionweb.com/de}, and DeepAI\footnote {https://deepai.org/3d-model-generator} allow the creation of visualizations based on prompts, which can be in the form of a textual or auditory description, or even another visualization. GAI offers almost unlimited visualization possibilities, allowing the user to choose the content, type and style of the visualization. This makes GAI particularly promising for exposure visualization in the context of CPTSD. Individual memories that are key to effective trauma processing, and that would not be visualizable without great financial or time expense, can now be visualized within minutes. Subsequently, the patient is given the opportunity to adjust the visualization as often as needed. Previously, recall was usually done before the visualization process because visualization was too costly to include in the early stages of traumatic memory recall. The flexibility gained is especially beneficial for the exploratory recall process of CPTSD~\cite{vanderkolkDissociationFragmentaryNature1995}. Visualizations in the early stages of the recall process can facilitate patient-therapist communication, help identify unclear memories, and can be used as an early indicator of whether a memory contains a traumatic stimulus. In addition, the patient can choose the level of detail of the visualization, which allows for a more patient-centered level of exposure and reduces the risk of unintentionally intense trauma flashbacks (\textbf{R2}). The early feedback and more customizable level of exposure could, in turn, have a positive impact on the overall effectiveness of the therapy.

\section{Challenges \& Risks of GAI-Based Visualization}
Although the use of GAI in exposure therapy for CPTSD appears to solve a number of critical challenges, it also creates challenges and puts patient safety at risk. GAI is often associated with a black box with uncertain, only partially controllable output~\cite{hassijaInterpretingBlackBoxModels2024}. Due to the limited accessibility of visualization-based exposure for CPTSD, there is little knowledge on how exposure visualizations should be designed to be effective. An ongoing feasibility study by the authors investigates the perception and acceptance of individualized VR exposure scenarios for CPTSD (DRKS00032739\footnote{registered at the German Clinical Trials Registry: https://drks.de/search/en}). Early findings suggest that traumatic memories often contain specific aspects that are responsible for their recall. This suggests that the extent of exposure scenarios could be significantly reduced, which is in line with previous work~\cite{bohus2024dialektisch}. However, we could not accurately predict which aspects would be responsible for the recall. 
Consequently, the traumatic memory recall process and the GAI-based visualization are exploratory in nature. This leads to an extraordinarily high degree of uncertainty about the course of the exposure visualization, which is further increased by the GAI, which is still prone to artifacts and inconsistencies in the visualizations. Regarding the risk of unintended recall (\textbf{R2}) and the risk of retraumatization (\textbf{R4}), it is clear that the feasibility of this concept depends on a thorough understanding of effective exposure design and the challenges and risks associated with GAI to minimize uncertainty and increase patient safety. The goal of exposure therapy is a controllable confrontation in a controlled setting. A potential system solution should support this at best, but should not interfere with it at least.  

\noindent A further challenge is the limited ability of modern GAI systems to make detailed edits. A small adjustment to the prompt, or even a request to adjust the previous visualization, can result in a completely different visualization.
This arguably poses a challenge to the iterative elaboration of a traumatic memory visualization. Patients may become frustrated if the requested adjustments undo previous progress. This could reduce acceptance of the system and increase the risk of dropout (\textbf{R5}). Uncontrolled change also increases the risk of unintended trauma flashbacks. Therefore, we need to understand how to minimize unintentional editing of visualizations and what safety mechanisms could be incorporated to further increase safety. Apparent approaches could be the prior examination of the visualizations by the therapist.

\subsection{Ethical Concerns and Potential Risks}
General criticism of generative AI concerns the bias of content due to the AI being trained on data from wealthy communities~\cite{bender2021dangers,Nyariro2022Integrating,kenthapadi2023GenerativeAI,bosco2025Black}. Hence, visualizations might be biased towards traditions, culture, and appearances of these communities. Another concern is the gender bias in GAI~\cite{geoffrey2024GenderBias}. Regarding exposure visualization, both cultural and gender bias could not only cause disengagement by patients, as they do not feel understood~\cite{randazzo2023TraumaInformedDesign}, but exposure could also remain ineffective due to the inability to visualize specific appearances not included in the training data. However, this is likely to concern a considerable proportion of the target group. Research found that people of non-binary or transgender have a three in ten times higher prevalence of PTSD than the general community and that the risk of PTSD may vary depending on the sociocultural background~\cite{brewin2025post}. Hence, these limitations should be cautiously explored to comprehend the potential effects of GAI on the acceptance and efficacy of trauma therapy and identify feasible countermeasures.\\ 

Another core concern is the deliberate \emph{circumvention of restraints and policies} for harmful content. Arguably, in some cases, especially considering childhood trauma with physical abuse, such circumvention may be necessary to visualize key traumatic memories that enable successful therapy. However, as there is little knowledge on essential aspects of exposure visualizations for CPTSD, it is unclear whether there are more harmless alternatives that could also be effective for such use cases. The choice of less harmful exposure scenarios could reduce the risk of hyperarousal and retraumatization~\cite{fordTraumaMemoryProcessing2018,duckworth2012retraumatization}. However, the choice of harmful content could also lead to a more detailed processing of the traumatic memories and hence could lead to a better therapy outcome~\cite{rizzo2021fromcombat}.\\

A further critical aspect of GAI-based exposure visualization is the distortion of reality. It is an already known risk of traumatic memory retrieval in general. Limited access to one's own memories can lead patients to mistakenly believe that they remember something~\cite{mcnally2005remembering}. However, this effect can also be used for good. Imagery rescripting targets deliberate memory distortion to positively reframe traumatic memories~\cite{HOLMES2007297}. Therefore, therapists need to reflect on the formulation of questions about traumatic memories and the respective exposure designs to control potential distortion effects and ensure the patient's safety. Reality distortion is also frequently discussed in HCI~\cite{tseng2022the,bonnail20232023,slater2020ethics}. Research has shown that memories of real and virtual experiences can be manipulated by the design of virtual experiences. Arguably, this could be used as a dark pattern to manipulate the user. 
GAI-based exposure visualization could put patients suffering from CPTSD at such risk. The lack of control of generative AI could also cause this false memory effect (\textbf{R6}). This concerns the patient-AI communication in the context of prompting and the visualizations themselves. Hence, feasibility depends on reflected design with full control by the therapist in order to intervene in case of potential harmful distortion. Research is needed to understand the extent of this risk and how therapist-sided control may be realized to enable controlled exposure.\\

There is a need for experience-based guidelines and policies tailored to the specific use case of GAI-based exposure visualization for CPTSD. Policymakers and ethicists need to develop guidelines that take into account the specific CPTSD-related risks of exposure to innocuous and harmful visualizations. In addition, we recommend close collaboration with people with CPTSD when developing guidelines. This is critical for accessibility, as language barriers may limit comprehension. A lack of understanding of the functionality of AI may, in turn, reduce trust and acceptance of the use of GAI in trauma therapy (\textbf{R5}).

\section{Considerations to Promote Feasibility}
HCI has explored various ways to improve the quality of health care by identifying current shortcomings and needs for patient-centered design. For example, the guidelines of trauma-informed conceptualizations from other research fields~\cite{huang2014samhsa} have been transferred to HCI to provide guidelines for system design to increase accessibility for traumatized individuals~\cite{randazzo2023TraumaInformedDesign}. To minimize potential harm, they provide recommendations for creating a safe space that promotes trust through transparency, empowers the individual user, recognizes their abilities, and facilitates peer interaction. They also address cultural, historical, and gender issues.
It is unlikely that these guidelines can be fully addressed. For example, the goal of a system where all users feel safe is recommended by providing a physically safe space with social interaction that promotes safety. However, due to its intended use (\textbf{R1}), patients are unlikely to perceive the system as a safe space, even if the social interaction and environment are designed to promote safety~\cite{rainey1987laboratory}. Nevertheless, we summarize some aspects of trauma-informed design that might be useful to examine for their ability to promote system accessibility.\\

\noindent When it comes to creating a safe and trustworthy space, the design of patient, therapist, and system interactions is critical. For example, trust can be further enhanced if the patient-GAI interaction undergoes a conscious design of moral agency. Wester et al. recommend that the system should be empathic and use positive, reassuring wording, but tailored to the communication style of the target audience~\cite{wester2024ChatbotMoralAgency}. On the contrary, other work, such as that of Shao~\cite{shao2023Empathetic}, showed an increase in disclosure for introducing empathy to chatbots in health care. However, since a therapist necessarily accompanies a GAI-based visualization, these effects are not expected.\\ The severity, type, and onset of traumatization may influence the patient's communication style. In addition, the accessibility of traumatic memories~\cite{mcnallyDispellingConfusionTraumatic2007} may influence communication style and should be considered~\cite{herman1987recovery}. This is important not only to support perceived safety and trust, but also to improve communication for effective collaborative exploration of amnesic memories. Previous work has emphasized that the use of mainstream language can inhibit the sharing of experiences for traumatized individuals~\cite{Ardener1975-ARDPW}. A system that adapts to the user's communication style and summarizes the patient's descriptions could promote the accessibility of shared memories and help the patient find a formulation~\cite{randazzoIfSomeoneDownvoted2023}. It has also been shown to be beneficial to target the creation of a narrative for successful memory retrieval~\cite{vanderkolkDissociationFragmentaryNature1995,raeder2023narrative}. However, the system should avoid suggesting in which direction the patient could continue the narrative to avoid a false memory effect (\textbf{R6}).
In terms of patient-therapist interaction, the mere presence of the therapist is important~\cite{vanderkolkDissociationFragmentaryNature1995}. Health care quality research emphasizes the need to integrate professional mental health support at all stages~\cite{chassinurgent1998,emanuelWhatExactlyPatient2009}. However, the design of the interaction should be well thought out, as the patient-therapist relationship is challenged by the dependency and attachment often associated with CPTSD~\cite{steeleDependencyTreatmentComplex2001}. It is also unclear to what extent attachment may also occur in the context of patient-GAI interaction. This should be investigated to ensure a safe design solution.

Arguably, the main challenge in designing a transparent and trustworthy system will be to minimize the black box effect of GAI~\cite{deChoudhuryIntegratingAHI2018}. AI should be carefully implemented with therapists involved in the entire implementation process to minimize the black box effect associated with intelligent systems~\cite{deChoudhuryIntegratingAHI2018}. As individualized exposure therapy has been shown to be most effective for the target group, not only concerning the exposure design but also the communication and pace within the therapy, the system should restrict the actual therapy design as little as possible. The inclusion of natural intelligence in the form of human feedback is critical to success~\cite{deChoudhuryIntegratingAHI2018}. The experience of therapists is needed to gather extensive experience about the use of GAI in such a context. Developers should implement features that prevent the unintentional triggering of flashbacks (\textbf{R2}). The therapist should have full control over the level of exposure and be able to adjust the communication style of the system. Another promising solution is provided by Explainable AI, which aims to increase the transparency of the functionality and decisions made by the AI in order to increase the trust of the users~\cite{alotaibi2024AIMental}. With the use of explainable AI, not only could patients more effectively communicate their needs and describe their memories, but the increased transparency is also likely to reduce the fear of unintended recall and consequently increase trust. However, research is needed to investigate this. Adapting the communication style of chatbots in the healthcare industry has already been shown to be beneficial.
   
\section{Conclusion}
The integration of generative artificial intelligence (GAI) in trauma exposure therapy presents both transformative opportunities and critical challenges, particularly for individuals with Complex PTSD. While GAI enables unprecedented flexibility in creating individualized exposure visualizations, its inherent uncertainties, ethical concerns, and risks associated with retraumatization and unintended memory distortion necessitate careful consideration. The exploratory nature of both CPTSD-related memory retrieval and AI-based content generation underscores the need for a structured, therapist-guided design process to ensure safety and efficacy.

Future research must focus on the exploration of patient-therapist-GAI interaction, as a more profound comprehension of risks and design-related opportunities is needed to achieve a feasible solution that can be used in trauma exposure therapy for CPTSD. In compliance with existing guidelines for the design of technology in health care, we argue that designs should focus on therapist control and transparency to minimize unintended recall and foster trust through safety mechanisms. This should be included in all stages of the GAI-based exposure therapy process and on the side of all three stakeholders: patient, therapist, and system. However, we want to point out that the unconditional prioritization of trauma-informed design principles and ethical and regulatory frameworks may interfere with the efficacy and acceptance of the design solution. Safe and effective recall of traumatic memories is unlikely to be possible, given the intense emotional and physical reactions involved. To some extent, safety may even counteract efficacy. The feasibility of GAI-based exposure crucially depends on an experience-based design that optimizes the balance of efficacy and acceptance on the one hand and safety, ethical, and regulatory concerns on the other hand. Provided its implementation remains patient-centered and guided by clinical expertise, GAI has the potential to significantly improve access to and effectiveness of trauma therapy for CPTSD.

\bibliographystyle{ACM-Reference-Format}

\bibliography{bibliography}

\end{document}